\documentclass[11pt]{article}

\usepackage{times}
\usepackage{bbm}
\usepackage{float}
\usepackage{amsmath,amsfonts}
\usepackage{hyperref}
\usepackage{graphicx}
\usepackage{subfigure}
\usepackage{multirow}
\usepackage{blkarray}
\usepackage{tabularx}
\usepackage{adjustbox}
\usepackage{booktabs}
\usepackage{array}
\usepackage{amsthm,amscd,amsxtra,amsfonts,amsmath,amssymb,multirow}
\usepackage{algorithm,algorithmic}
\usepackage{epstopdf}
\usepackage{xcolor}
\usepackage{setspace}
\hypersetup{
	colorlinks=true,
	linkcolor=blue,      
	urlcolor=magenta,
	citecolor=blue,
}

\theoremstyle{definition}

\theoremstyle{remark}

\theoremstyle{example}

\theoremstyle{conjecture}

\title{Benchmarking AlphaFold3's protein-protein complex accuracy and machine learning prediction reliability for binding free energy changes upon mutation}

\author
{JunJie Wee$^{1}$, and Guo-Wei Wei$^{1,2,3,\ast}$\\
\normalsize{$^{1}$Department of Mathematics, Michigan State University, East Lansing, MI 48824, USA}\\
\normalsize{$^{2}$Department of Biochemistry and Molecular Biology,}\\ 
\normalsize{Michigan State University, East Lansing, MI 48824, USA}\\
\normalsize{$^{3}$Department of Electrical and Computer Engineering,} \\
\normalsize{Michigan State University, East Lansing, MI 48824, USA}\\
\normalsize{$^\ast$ Address correspondences to Guo-Wei Wei. E-mail: weig@msu.edu}
}

\date{}

\begin{document}


\baselineskip24pt


\maketitle

\begin{abstract}
AlphaFold 3 (AF3), the latest version of protein structure prediction software, goes beyond its predecessors by predicting protein-protein complexes. It could revolutionize drug discovery and protein engineering, marking a major step towards comprehensive, automated protein structure prediction. However, independent validation of AF3's predictions is necessary. Evaluated using the SKEMPI 2.0 database which involves 317 protein-protein complexes and 8338 mutations, AF3 complex structures give rise to a very good Pearson correlation coefficient of 0.86 for predicting protein-protein binding free energy changes upon mutation, slightly less than the 0.88 achieved earlier with the Protein Data Bank (PDB) structures. Nonetheless, AF3 complex structures led to a 8.6\% increase in the prediction RMSE compared to original PDB complex structures. Additionally, some of AF3's complex structures have large errors, which were not captured in its ipTM performance metric. Finally, it is found that AF3's complex structures are not reliable for intrinsically flexible regions or domains. 

\end{abstract}

\newpage
\begin{spacing}{0.1}
\tableofcontents
\end{spacing}

\newpage

\section{Introduction}

AlphaFold3 (AF3) is the latest iteration of the groundbreaking protein structure prediction software developed by Google DeepMind and Isomorphic Labs\cite{abramson2024accurate}. Building on the success of its predecessors, AF3 brings significant advancements to the field of computational biology. One of the key features of AF3 is its ability to predict protein-molecule complexes, including those involving DNA and RNA. This represents a major leap from AF2, which primarily focused on predicting the three-dimensional (3D) structures of proteins. The remarkable success of AF3 has opened up new possibilities for extending its purpose beyond fundamental structural prediction tasks to include protein-molecule complexes.

The significance of accurately predicting 3D protein structures using AlphaFold cannot be overemphasized. Fundamentally, 3D protein structures are instrumental in revealing protein functions, comprehending various biological interactions, and designing drugs. Numerous prevalent diseases, including Alzheimer’s and Parkinson’s, are linked to aberrant protein structures. Consequently, the automated prediction of protein folds from sequences has emerged as a crucial challenge in biology and is often referred to as the holy grail of molecular biophysics. In 2019, AlphaFold’s \cite{jumper2021highly} leading performance in predicting 25 structures out of 43 test proteins sparked enthusiasm among researchers about the potential future of AI-based protein structure prediction. For instance, DeepFragLib by Wang et al. \cite{wang2019improved} signifies a novel progression in ab initio protein structure prediction. Thereafter, the launch of AlphaFold ignited a transformative shift in the way we model protein structures and their interactions \cite{jumper2021highly}. AlphaFold also opened up a vast array of possibilities in protein folding, protein engineering, and design \cite{wei2019protein,qiu2023persistent,lin2023evolutionary,mirdita2022colabfold,tunyasuvunakool2021highly,dejnirattisai2022sars,nunes2023alphafold2}.

Data-driven machine learning models have demonstrated great power by utilizing 3D protein-protein complexes. Protein-protein interactions (PPIs) also play a significant role in nearly all cellular and biological activities. In the study of PPIs, mutation-induced effects play a paramount role in evolutionary biology, cancer biology, immunology, directed evolution, and protein engineering. Data-driven machine learning models have targeted the study of mutation-induced effects on protein stability and PPI binding affinities. Computational approaches have primarily served as a viable alternative to experimental mutagenesis methods. Previously, scientific communities have naturally extended the capabilities of AlphaFold by expanding the protein structural database\cite{varadi2022alphafold}. With the accessibility of AF3 through the AlphaFold Server, AF3 can potentially become a valuable tool for advancing deep learning models towards applications of PPIs. 

Nonetheless, it was found that the prediction accuracy of 
protein engineering, which iteratively optimizes protein fitness
by screening the gigantic mutational space, drops when AlphaFold2 structures 
were used \cite{qiu2023persistent}. Therefore, it is imperative to independently validate AF3's accuracy and reliability for PPI analysis and mutation-induced protein-protein binding free energy (BFE) change predictions.

One of the very successful approaches for mutation-induced protein-protein BFE change predictions is the topology-based network tree (TopNetTree) \cite{wang2020topology}, which is based on topological deep learning (TDL), introduced in 2017 \cite{cang2017topologynet}. TDL is an emerging paradigm in machine learning, based on topological data analysis (TDA) \cite{edelsbrunner2008persistent,zomorodian2004computing}. TDA, a branch of mathematics, focuses on understanding the shape and structure of data and is exceptionally successful in improving standard approaches by contributing novel topological characterization. TDL captures protein structures, simplifies the structural complexity of biomolecules, and embeds physical interactions into topological invariants. 

TDA has had tremendous success through TopNetTree \cite{wang2020topology}, PerSpect-EL \cite{wee2022persistent}, HCML \cite{liu2022hom}, and TopNetmAb \cite{chen2022persistent} by incorporating topological fingerprints from persistent homology (PH) and persistent Laplacian to predict PPI BFE changes upon mutation and subsequently, BFE changes due to mutations in the SARS-CoV-2 receptor binding domain (RBD) - Angiotensin-converting enzyme 2 (ACE2) complexes. In particular, persistent Laplacian provides both the topological information in PH and the homotopic shape of evolution \cite{wang2020persistent}. Using persistent Laplacian, Omicron BA.4 and BA.5 were predicted to be new dominant variants two months before the World Health Organization (WHO) made the announcement \cite{chen2022persistent}.  Persistent Laplacian is also incorporated in the recent integration between TDL and pre-trained ESM transformer features for predicting mutation-induced protein solubility changes, which establishes the state-of-the-art method for protein-protein BFE change predictions \cite{wee2024integration}.


In this work, we evaluate AF3's complex structure accuracy and mutation-induced BFE change prediction reliability using the largest PPI database, SKEMPI 2.0, which involves 317 protein-protein complexes and 8,330 mutation-induced BFE changes \cite{jankauskaite2019skempi,wang2020topology}. First, we consider AF3's PPI structure accuracy. Additionally, we examine AF3's reliability in predicting protein-protein BFE changes upon mutation using the most advanced TDL method \cite{wee2024integration}.  
Our results demonstrate that AF3-predicted complexes achieved a relatively good Pearson correlation coefficient of 0.86, which is slightly less than the 0.88 reported earlier \cite{wee2024integration}. However, we found that AF3 results in 8.6\% increase of root-mean-square error (RMSE) compared to the original complexes in the Protein Data Bank (PDB) for BFE change prediction. Through our analysis, we also discovered that some structurally misaligned AF3 protein-protein complexes are not captured by AF3's ipTM performance metric. Finally, 
AF3 complex predictions may not be reliable for highly flexible protein domains. 
 

\section{Results}

In this section, we assess the accuracy and reliability of AF3 predictions of PPI complexes. Here, we use the accessible AlphaFold Server of AF3 to predict the protein-protein complexes from SKEMPI 2.0 database \cite{jankauskaite2019skempi}. This database contains the S8338 dataset for mutation-induced BFE changes, which is one of the most comprehensive datasets collected on how mutations can alter the binding affinity in PPIs, with over 8,338 entries of single mutations. However, SKEMPI 2.0 database involves certain PPI complexes, namely 3NVN and 4U6H,  with restricted viral pathogenic sequences in AlphaFold Server (accessed 11th May 2024). These complexes only make up 8 samples in the S8338 dataset. Therefore, we incorporate the remaining 8330 single mutation-based samples generated from 317 PPI complexes. The 317 predicted AF3 complexes are utilized to predict the 8330 mutation-induced binding free energy changes. More details on the SKEMPI 2.0 database can be found in the Supplementary Information.

\subsection{Validation Performance} 
In the validation test, we conduct a 10-fold cross-validation on MT-TopLap$_{\rm AF3}$ by predicting the mutation-induced BFE changes using the features extracted from the remaining 317 AF3 structures. Existing topology-based models like TopLapNetGBT and TopNetTree, which have proven to be successful, have undergone training and validation using the S8338 dataset \cite{chen2022persistent,wang2020topology}. 

\begin{figure}[!htbp]
	\centering 
	\includegraphics[width=\textwidth]{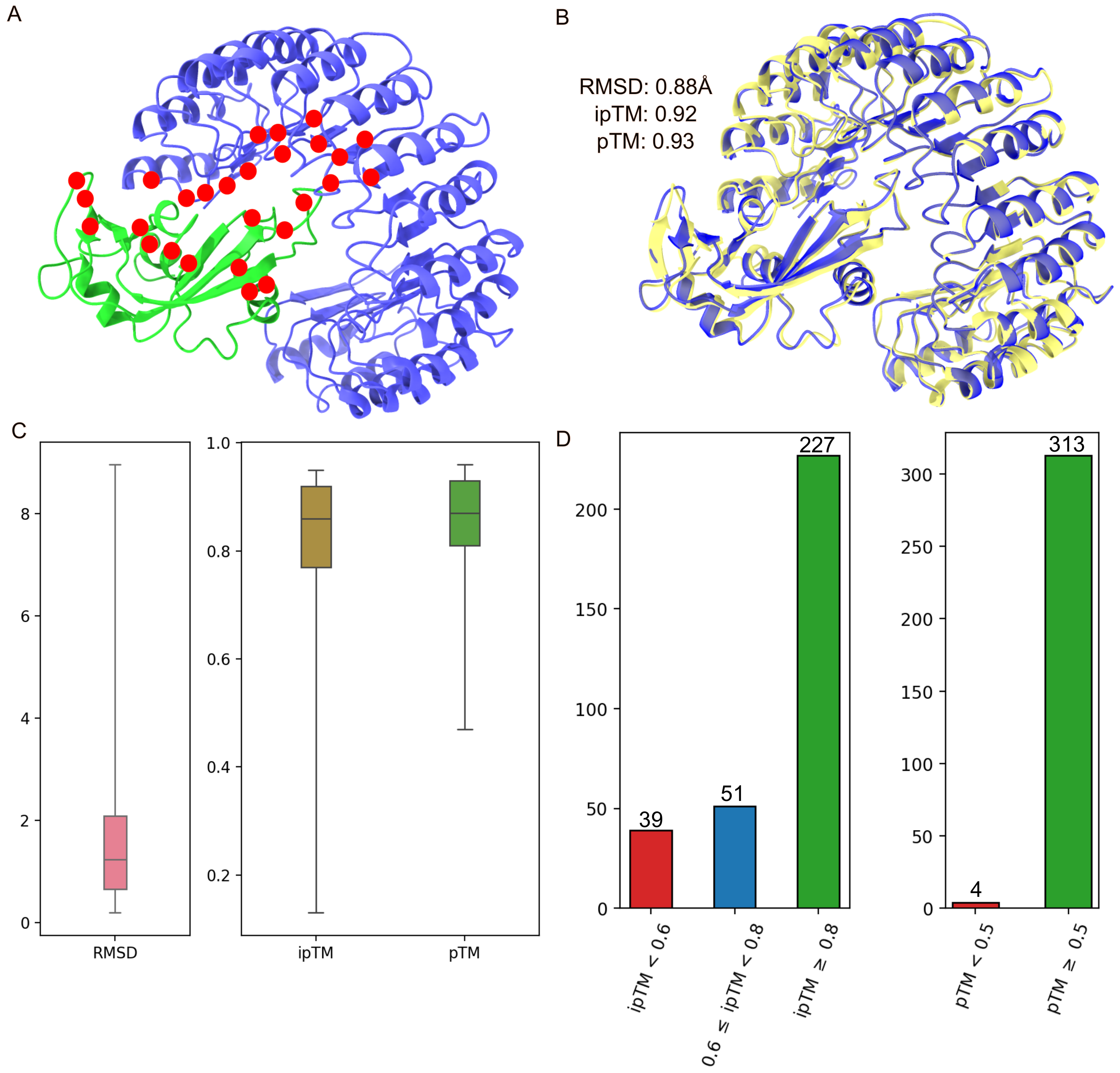}
	\caption{A: The cartoon representation of ribonuclease inhibitor-angiogenin complex (PDB ID: 1A4Y). The ribonuclease inhibitor is shown in blue and the angiogenin is in green. The mutation spots of 1A4Y in the S8338 dataset are indicated in red. B: The structural alignment of 1A4Y with its AF3 predicted complex. C: The boxplot for RMSD, ipTM and pTM distributions of 317 predicted AF3 protein-protein complexes. RMSDs refer to the overall RMSD calculated by structurally aligning an AF3 complex with its original PDB complex. D: The breakdown of AF3 protein-protein complexes based on their ipTM and pTM scoring criteria.}
	\label{fig:1A4Y}
\end{figure}
Table \ref{tab:S8338} shows the 10-fold cross validation results of MT-TopLap$_{\rm AF3}$ models for the BFE change prediction for mutations on the S8338 dataset. The performance of MT-TopLap$_{\rm AF3}$ model is evaluated based on the Pearson correlation coefficient ($R_p$) and RMSE. Their definitions are included in the Supplementary Information. MT-TopLap$_{\rm AF3}$ achieved an $R_p$ of 0.86 and RMSE of 1.025 kcal/mol. In comparison to non-topology-based models such as mCSM-PPI2 \cite{rodrigues2019mcsm}, MT-TopLap$_{\rm AF3}$ displayed an 18\% improvement in its RMSE. 

\begin{table}[H]
	\centering
	\begin{tabular}{@{}lc||lc@{}}
		\toprule
		Method           & $R_p$                     & Method               & $R_p$                \\ \midrule
		MT-TopLap     & \multicolumn{1}{c||}{0.88} & mCSM-PPI2\cite{rodrigues2019mcsm}             & 0.82                 \\
		TopLapNetGBT\cite{chen2022persistent}     & \multicolumn{1}{c||}{0.87} & LapNet\cite{chen2022persistent}                & 0.81                 \\
		TopLapNet\cite{chen2022persistent}         & \multicolumn{1}{c||}{0.87} & LapGBT\cite{chen2022persistent}                & 0.80                 \\
		TopNetGBT\cite{chen2022persistent}         & \multicolumn{1}{c||}{0.87} & \multicolumn{1}{c}{} &                      \\
		TopNet\cite{chen2022persistent}            & \multicolumn{1}{c||}{0.86} & \multicolumn{1}{c}{} &                      \\
		MT-TopLap$_{\rm AF3}$  & \multicolumn{1}{c||}{0.86$^*$} & \multicolumn{1}{c}{} &                      \\
		TopLapGBT\cite{chen2022persistent}         & \multicolumn{1}{c||}{0.85} & \multicolumn{1}{c}{} &                      \\
		TopGBT\cite{chen2022persistent}            & \multicolumn{1}{c||}{0.85} & \multicolumn{1}{c}{} &                      \\
		LapNetGBT\cite{chen2022persistent}         & \multicolumn{1}{c||}{0.83} & \multicolumn{1}{c}{} &                      \\ \bottomrule
	\end{tabular}
	\caption{Comparison of the Pearson correlation coefficients ($R_p$) of MT-TopLap, MT-TopLap$_{\rm AF3}$ and existing state-of-the-art methods for 10-fold cross-validation of  the S8338 dataset. Results of existing state-of-the-art methods are obtained from \cite{chen2022persistent,rodrigues2019mcsm}. *Note that MT-TopLap$_{\rm AF3}$ is validated with 8330 samples as PDB ID: 3NVN and 4U6H cannot be predicted by using AlphaFold server (accessed 11th May 2024) as they are detected as ``restricted sequences from a small number of viral pathogens."}
	\label{tab:S8338}
\end{table}

\begin{figure}[!htbp]
	\centering 
	\includegraphics[width=.9\textwidth]{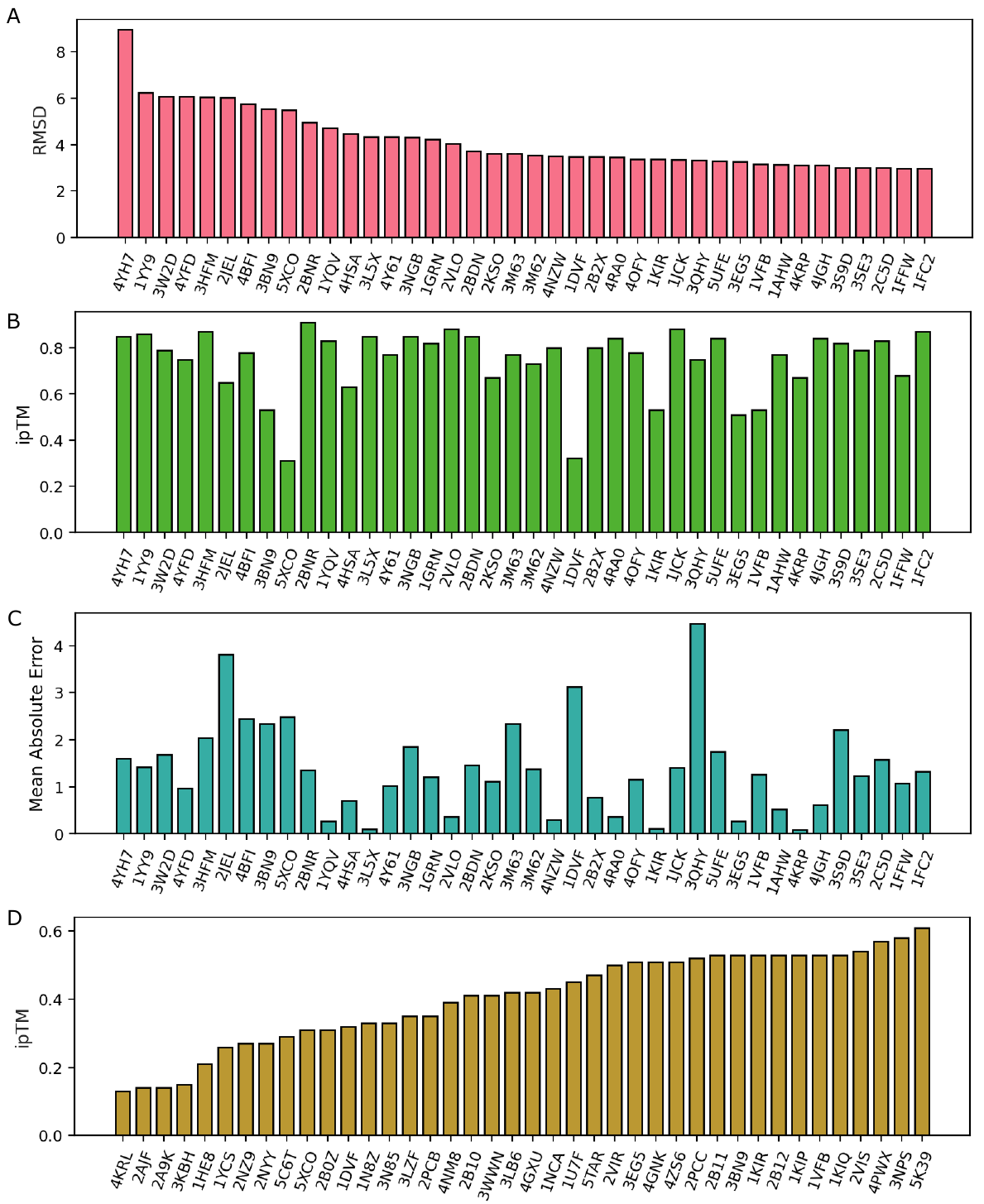}
	\caption{A: Top 40 protein-protein complexes with poorest RMSD alignment scores. B: The ipTM scores for protein-protein complexes in A. C: The mean absolute prediction error for mutation-induced binding free energy changes of the protein-protein complexes in A. D: Top 40 protein-protein complexes with ipTM scores below 0.6.}
	\label{fig:ranking}
\end{figure}

\subsection{Structural Alignment Performance}
Using with the original PDB structures for the S8338 dataset, the original MT-TopLap achieved an $R_p$ of 0.88 and RMSE of 0.937 kcal/mol. This implies an increase in 8.6\% of RMSE due to the AF3 predicted complex structures. To investigate further, we analyze the AF3 complexes by ranking them based on their structural alignment RMSD, ipTM, and pTM scores. RMSD scores are obtained by superimposing AF3 complexes with original PDB complexes. The ipTM score, which was first introduced in AlphaFold-Multimer, is a metric that evaluates the accuracy of the predicted interface in a protein-protein complex \cite{evans2021protein}. For example, an ipTM score above 0.8 indicates that the PPI complex generated is highly confident prediction. On the other hand, an ipTM score below 0.6 suggests that the predicted structure is likely to be incorrect while any ipTM score between 0.6 and 0.8 includes prediction that are correct or wrong \cite{evans2021protein}. 
Another AF3 scoring is pTM which is a comprehensive measure of the accuracy of prediction of the entire structure of the complex. It represents the predicted TM score for a superposition between the predicted structure and the assumed true structure. However, a good pTM score above 0.5 may be due to a correct prediction of the larger protein dominated in the protein-protein complex although the other partner protein has a poor prediction.

Figure \ref{fig:1A4Y} shows the statistics of 317 AF3 structures from S8338 dataset. The average RMSD, average ipTM and average pTM calculated are $1.61$\AA, 0.803 and 0.847 respectively. In Figure \ref{fig:1A4Y}A, complexes that are considered outliers have a poor RMSD greater than 4\AA. Despite this, 71.6\% of the complexes have an high ipTM score of at least 0.8, and 98.7\% of the complexes have a pTM score of at least 0.5. Essentially, this shows that most of the AF3 predicted complexes have high ipTM but also have low RMSD. The four complexes (i.e. 1DVF, 3LB6, 3LZF and 5TAR) with pTM below 0.5 also have ipTM scores below 0.6.

\section{Discussion}

To analyze which AF3 PPI complex is not well predicted, we first rank the 317 AF3 complexes with their RMSD and ipTM scores. Figure \ref{fig:ranking}A and B shows the top 40 PPI complexes according to its RMSD (from highest to lowest) and ipTM (from lowest to highest). Note that lower RMSD and higher ipTM indicates better performance. Clearly, a poor ipTM score does not necessarily translate into poor RMSD performance. Comparing Figures \ref{fig:ranking}A and B, only six complexes, i.e., 5XCO, 1DVF, 3EG5, 1KIR, 3BN9, 1VFB, have both low ipTM and low RMSD scores. Similarly, the same six complexes appear in both Figures \ref{fig:ranking}A and D. Therefore, there is little correlation between ipTM score and prediction RMSD. In general, RMSD also influences the final absolute prediction errors in Figure \ref{fig:ranking}C. Future improvements to AF3 should incorporate RMSD as part of its prediction performance.

\begin{figure}[!htbp]
	\centering 
	\includegraphics[width=\textwidth]{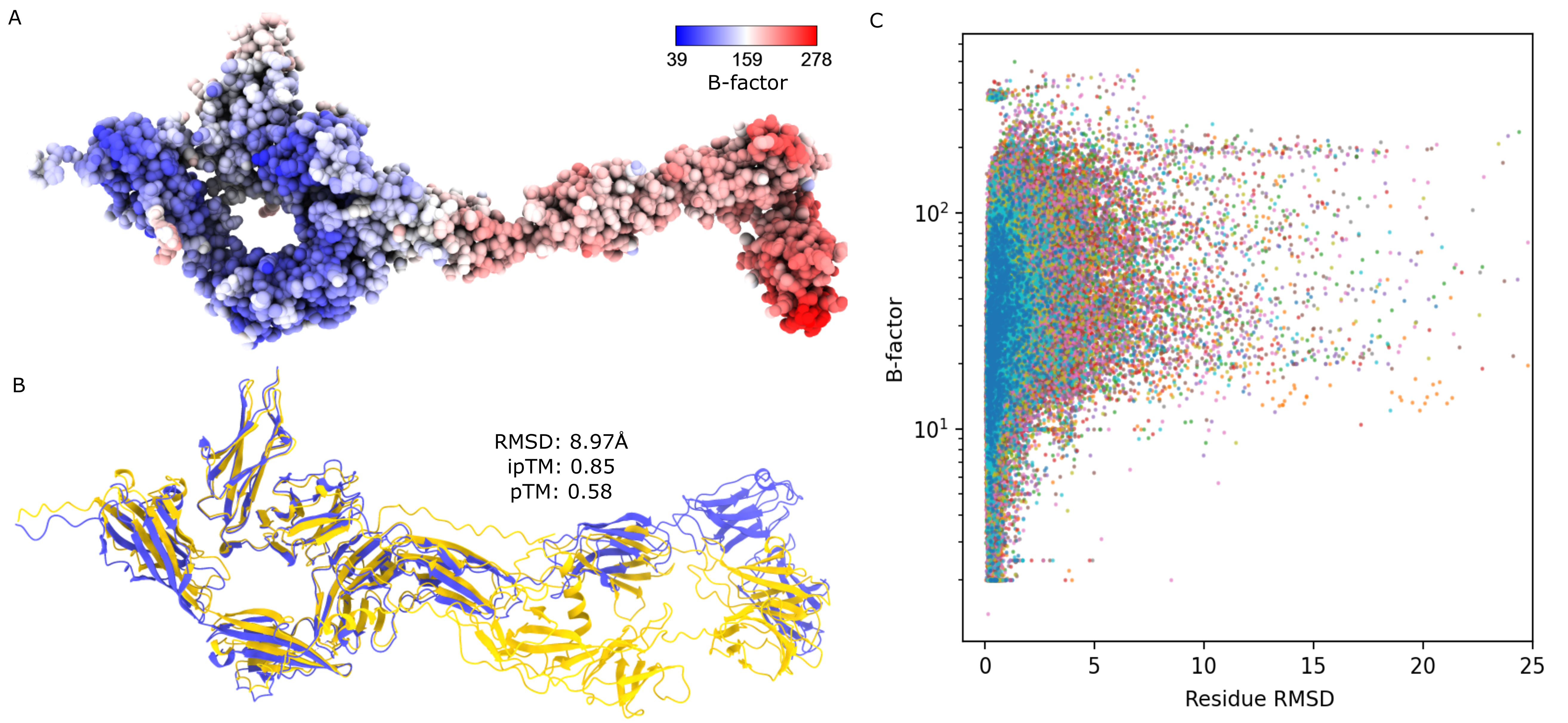}
	\caption{A: The B-factor representation of PDB ID: 4YH7. Atoms colored in red have high B-factors while atoms colored in blue have lower B-factors. B: Illustration of structural alignment for PDB ID: 4YH7 (in blue) and its AF3 predicted complex (in orange). C: The residue-based B-factor values plotted against all the residue RMSD calculated from all 317 AF3 predicted complexes. A $\log$-scale is taken on the $y$-axis.}
	\label{fig:4YH7}
\end{figure}

Another potential limitation of AF3 is illustrated in Figure \ref{fig:4YH7}A and B. Figure \ref{fig:4YH7}A shows the PPI complex of PDB ID: 4YH7 colored by B-factor values. High B-factors or high flexible protein regions are colored in red while low B-factors or more rigid protein regions are colored in blue. Figure \ref{fig:4YH7}B shows the structural alignment of PDB ID: 4YH7 (in blue) with its AF3 predicted complex (in orange). It can be observed that the regions in Figure \ref{fig:4YH7}A with high B-factors correspond to the region in Figure \ref{fig:4YH7}B where the residues are misaligned between AF3 and PDDB structures. 
Here, the RMSD per residue is calculated based on the backbone atoms of the protein. To investigate this further, we observe a similar pattern when all residue-based B-factors are plotted against residue-based RMSD for all 317 PPI complexes in Figure \ref{fig:4YH7}C. The colors in Figure \ref{fig:4YH7}C represent the 20 canonical amino acid types. In general, Figure \ref{fig:4YH7}C shows a huge amount of residues with high B-factors and high RMSD. 
Based on these analyses, it is concluded that AF3 complex prediction is not reliable for intrinsically flexible regions or domains of  PPI complexes.


\section{Conclusion}
We examine the use of AlphaFold3 (AF3) predicted complex structures for the prediction of protein-protein binding free energy (BFE) changes upon mutation for the S8338 dataset in the SKEMPI 2.0 database. We noted that AF3 complex structures lead to a relatively good predicted Pearson correlation coefficient of 0.86  but have generated 8.6\% more in RMSE compared to using PDB complex structures. 
Essentially, AF3 is a promising technology for predicting protein-protein complexes, even capable of contributing to mutation-induced BFE change predictions. 

In total, 317 AF3 complexes inferred by the SKEMPI 2.0 database were structurally aligned with original ones in the PDB to investigate AF3 predictions that have poor RMSD in structural alignment. Through our analysis, we discovered that current AF3 performance metrics such as ipTM does not correlate with RMSD alignment scores. On the other hand, preliminary findings show that high RMSD are strongly correlated to high B-factors, indicating AF3 predictions are not reliable for highly flexible protein regions or domains.   

\section{Methods}
\subsection{Feature generation for MT-TopLap}

Among all features used in MT-TopLap prediction, persistent Laplacians (PL) plays the most significant role in validating the reliability of AF3 structures. With accurate 3D SARS-CoV-2 receptor binding domain (RBD) - Angiotensin-converting enzyme 2 (ACE2) complexes, PLs correctly predicted the dominance of the Omicron BA.4 and BA.5 two months preceding the public announcement by World Health Organization (WHO) in June 2022 \cite{chen2022persistent}.  

In this section, we provide the important main mathematical framework instrumental for understanding element-specific PL descriptors applied in our feature generation. Element-specific topological approaches were introduced in earlier work\cite{cang2017topologynet, cang2017analysis}. In the simplicial complex and PL framework \cite{wang2020persistent}, we emphasize their importance in identifying both harmonic and non-harmonic spectral characteristics that are crucial for understanding PPIs. 

To generate the PL features for our MT-TopLap models, we categorize the atoms in PPIs into several subsets. These subsets are 
\begin{itemize}
	\item Atoms in mutation sites $\mathcal{A}_{m}$,
	\item Atoms in mutation neighborhood, i.e. Atoms within a distance $r$ from the mutation site $\mathcal{A}_{mn}(r)$,
	\item Protein 1's atoms within distance $r$ of the binding site  $\mathcal{A}_{\text{P}_1}(r)$,
	\item Protein 2's atoms within distance $r$ of the binding site $\mathcal{A}_{\text{P}_2}(r)$.
\end{itemize}
Additionally, we also group atoms into different element specific categories such as $\{\text{C, N, O}\}$, denoted as $\mathcal{A}_{\text{ele}}$. These element-specific groups are crucial in identifying the various types of interactions in a PPI model, as per biophysical principles. For instance, the subsets $\mathcal{A}_{\text{C}}\cap \mathcal{A}_{\text{P}_1}(r)$ and $\mathcal{A}_{\text{C}}\cap \mathcal{A}_{\text{P}_2}(r)$ capture hydrophobic C-C PPIs while $\mathcal{A}_{\text{N}}\cap \mathcal{A}_{\text{P}_1}(r)$ and $\mathcal{A}_{\text{O}}\cap \mathcal{A}_{\text{P}_2}(r)$ capture hydrophillic N-O PPIs. 

In addition to groups specific to elements, we also utilize distance functions like ${\rm D}_{\text{mod}}$, which exclude interactions between atoms from the same subset. For interactions between atoms $A_i$ and $A_j$ in sets $\text{P}_1$ and $\text{P}_2$, ${\rm D}_{\text{mod}}$ is defined as follows:
\begin{equation}
{\rm D}_{\text{mod}}(A_i, A_j) = \begin{cases}
\infty, & \text{ if }A_i,A_j \in \text{P}_1 \text{ or } \text{ if }A_i,A_j \in \text{P}_2\\
{\rm DE}(A_i, A_j), & \text{otherwise}.
\end{cases}
\end{equation}
Here, ${\rm DE}(\cdot, \cdot)$ refers to the Euclidean distance between two atoms. 

By utilizing the groups specific to elements/sites, atoms are arranged into point clouds, which are then used to construct simplicial complexes. Specifically, a set of $k+1$ atoms from an element/site-specific subset forms $k+1$ independent points, which can be represented as a set $S=\{v_0, v_1, v_2, \cdots, v_k\}$. The convex hull of $k+1$ affinely independent points essentially forms a $k$-simplex. In simple terms, a point is a 0-simplex, an edge is a 1-simplex, a triangle is a 2-simplex, a tetrahedron is a 3-simplex, and so on for higher dimensions, forming a $k$-simplex. A simplicial complex is created by combining these finite simplices \cite{munkres2018elements,zomorodian2005topology,Edelsbrunner:2010,Mischaikow:2013}. There are numerous methods to construct a simplicial complex. For generating our persistent Laplacian-based features, we employed the Vietoris Rips complex for dimension 0 and the Alpha complex for dimensions 1 and 2. A Vietoris-Rips (VR) complex is an abstract simplicial complex that forms simplices by connecting any subset of points with a diameter not exceeding a certain threshold. On the other hand, an Alpha complex is a group of subcomplexes derived from a Delaunay triangulation, subject to a radius constraint that does not exceed a certain threshold. The Delaunay triangulation is a geometric structure that divides the convex hull of a set of points in a plane into triangles.

For a simplicial complex $K$, a $k$-th chain $c_k$ is the formal sum of $k$-simplicies in $K$, i.e. $c_k = \sum_i \alpha_i\sigma_i^k$. The $k$-th boundary operator $\partial_k: C_k \rightarrow C_{k-1}$ defined on a $k$-th chain $c_k$ is $\partial_k c_k = \sum_{i=0}^k \alpha_i\partial_{k}\sigma_i^k$. Here, the condition that the boundary of a boundary is empty must be satisfied. Defining the adjoint operator of $\partial_k$, i.e. $\partial_{k}^*:C_{k-1}\to C_{k}$, yields the inner product relation $\langle \partial_k(f), g \rangle=\langle f, \partial_{k}^*(g) \rangle,$ for every $f \in C_{k}$, $g \in C_{k-1}$. From here, the $k$-topological Laplacian, a linear operator $\Delta_k: C_k(K) \rightarrow C_k(K)$, computed as
$\partial_{k+1}\partial_{k+1}^* + \partial_{k}^*\partial_{k}$.

In matrix representations, we denote $\mathbf{B}_k$ as an $m\times n$ matrix of the boundary operators under the standard bases $\{\sigma_i^k\}^n_{i=1}$ and $\{\sigma_j^{k-1}\}^m_{j=1}$ of $C_k$ and $C_{k-1}$. In a similar way, the transpose boundary matrix $\mathbf{B}_k^\top$ is used to denote the matrix representation of $\partial_{k}^*$ with respect to the same ordered bases of the boundary operator $\partial_k$. This naturally leads to the $k$-combinatorial Laplacian matrix, which is an $n\times n$ matrix $\mathbf{L}_k$ computed as $ \mathbf{B}_{k+1}\mathbf{B}_{k+1}^\top + \mathbf{B}_k^\top \mathbf{B}_k$. For the special case $k=0$, $\mathbf{L}_0 = \mathbf{B}_1\mathbf{B}_1^\top$ as $\partial_{0}$ is understood as a zero map.

In our model, the key topological features are the eigenvalues of combinatorial Laplacian matrices. These eigenvalues are not dependent on the orientation choice \cite{horak2013spectra}. Moreover, the total number of zero eigenvalues, or their multiplicity, in $\mathbf{L}_k$ equates to the $k$th Betti number, $\beta_k$, as per the combinatorial Hodge theorem \cite{eckmann1944harmonische}. These Betti numbers are topological invariants that represent the $k$-dimensional holes in a $k$-simplicial complex. For instance, $\beta_0$, $\beta_1$, and $\beta_2$ denote the number of independent components, loops, and cavities, respectively. Overall, the zero and non-zero eigenvalues embody the harmonic and non-harmonic spectra of combinatorial Laplacian matrices. The non-harmonic spectra offer additional homotopic shape information that Betti numbers lack.

A single simplicial complex is not enough to capture all the topological information from a single protein-protein complex. By combining combinatorial Laplacian and multiscale filtration, we monitor the variations of harmonic and non-harmonic spectra by adjusting a filtration parameter such as radii/diameter for VR complex \cite{wang2020persistent}. For an oriented simplicial complex $K$, a filtration creates a nested sequence of simplicial complexes $(K_t)^m_{t=0}$ of $K$,
\begin{equation*}
\varnothing = K_0 \subseteq K_1 \subseteq \cdots \subseteq K_m=K.
\end{equation*} 
As the value of the filtration parameter increases, PL generates a sequence of simplicial complexes. Based on this nested sequence of simplicial complexes, we can produce a sequence of combinatorial Laplacian matrices $\mathbf{L}_{k}^0,\mathbf{L}_{k}^1,\mathbf{L}_{k}^2,\mathbf{L}_{k}^3\cdots,\mathbf{L}_{k}^n$ where $\mathbf{L}_{k}^t=\mathbf{L}_k(K_t)$. By changing the filtration parameter and performing diagonalization on the $k$-combinatorial Laplacian matrix, the characteristics of topology and spectrum can be examined from each $\mathbf{L}_k(K_t)$ ($0\leq t\leq n$). The eigenvalues of $\mathbf{L}_k(K_t)$ can be sorted in ascending order
$\text{Spectra}(\mathbf{L}_k^t) = \{(\lambda_1)_k^t, (\lambda_2)_k^t, \cdots, (\lambda_n)_k^t\}$ where $\mathbf{L}_k^t$ is an $n\times n$ matrix. The $p$-persistent $k$-combinatorial Laplacian can also be extended based on the boundary operator. Ultimately, these PL descriptors help MT-TopLap track the changes of harmonic and non-harmonic spectra in 3D protein-protein complexes, capturing both the intrinsic topological changes and homotopic shape evolution throughout its filtration process. Figure S2 depicts the harmonic and non-harmonic spectra generated from a point cloud using Alpha complex and ${\rm D_{\text{mod}}}$-based filtration. Further details of PL's mathematical framework is provided in Supplementary Information.

%
%

\section*{Data Availability}
The AF3 complexes can be generated using the AlphaFold Server and are readily available in \\
 \href{https://github.com/ExpectozJJ/MT-TopLap/alphafold/}{https://github.com/ExpectozJJ/MT-TopLap/alphafold/}. The original PDB files used in this study can be downloaded from the official Protein Databank: \href{https://www.rcsb.org/}{https://www.rcsb.org/}. The SKEMPI 2.0 database is also readily available from \href{https://life.bsc.es/pid/skempi2}{https://life.bsc.es/pid/skempi2}. 

\section*{Code Availability}
All source codes and models are publicly available at  
\href{https://github.com/ExpectozJJ/MT-TopLap/}{https://github.com/ExpectozJJ/MT-TopLap/}. A detailed set of instructions is available in the Supporting Information.

\section*{Supporting Information}
Supporting Information is available for supplementary tables, figures, and methods. 

\section*{Acknowledgments}
This work was supported in part by NIH grants  R01GM126189, R01AI164266, and R01AI146210, NSF grants DMS-2052983,  DMS-1761320, and IIS-1900473,  NASA grant 80NSSC21M0023,  MSU Research Foundation,  Bristol-Myers Squibb 65109, and Pfizer. 

\vspace{0.6cm}
\bibliographystyle{ieeetr}
\bibliography{refs}



\end{document}